\pdfoutput=0
\documentclass[aps,prl,reprint,showkeys,preprintnumbers]{revtex4-2}

\usepackage{epsfig}
\usepackage{amsfonts}
\usepackage{amsmath}
\usepackage{bm,bbm}
\allowdisplaybreaks[1]

\newcommand{\la}[1]{\label{#1}}
\newcommand{\be}{\begin{equation}}
\newcommand{\ee}{\end{equation}}
\newcommand{\ba}{\begin{eqnarray}}
\newcommand{\ea}{\end{eqnarray}}
\newcommand{\rmi}[1]{{\mbox{\scriptsize #1}}}
\newcommand{\fig}{Fig.~}

\newcommand{\eq}{Eq.~}
\newcommand{\eqs}{Eqs.~}

\newcommand{\nr}[1]{(\ref{#1})}

\newcommand{\nn}{\nonumber \\}
\newcommand{\fr}[2]{{\frac{#1}{#2}\,}}

\renewcommand{\vec}[1]{{\bf #1}}

\renewcommand{\eq}{eq.~}
\renewcommand{\eqs}{eqs.~}

\renewcommand{\fig}{fig.~}

\newcommand{\rmO}{{\mathcal{O}}}

\def\lsi{\raise0.3ex\hbox{$<$\kern-0.75em\raise-1.1ex\hbox{$\sim$}}}
\def\gsi{\raise0.3ex\hbox{$>$\kern-0.75em\raise-1.1ex\hbox{$\sim$}}}

\newcommand{\rmii}[1]{{\mbox{\tiny\rm{#1}}}}

\newcommand{\Tint}[1]{{\hbox{$\sum$}\!\!\!\!\!\!\!\int\,}_{\!\!\!\!\raise-0.9ex\hbox{$\scriptstyle{#1}$}}}
\newcommand{\Tinti}[1]{{{\Sigma}\!\!\!\!\raise0.3ex\hbox{$\int$}_\rmii{${#1}$}}}
\newcommand{\bi}{\begin{itemize}}
\newcommand{\ei}{\end{itemize}}
\newcommand{\hide}[1]{ }

\newcommand{\deltabar}{\raise-0.02em\hbox{$\bar{}$}\hspace*{-0.8mm}{\delta}}
\newcommand{\ddeltabar}{\raise-0.18em\hbox{$\bar{}$}\hspace*{-0.8mm}{\delta}}
\renewcommand{\P}{\mathcal{P}}

\newcommand{\taui}{\tau_{i}} 

\newcommand{\T}{\rmii{$T$}}

\newcommand{\mpl}{m_\rmii{pl}} 
\newcommand{\xiQ}{\varrho_\rmii{$Q$}} 
\newcommand{\hc}{\mbox{h.c.}}

\begin{document}

\title{Inflationary gravitational wave background as a tail effect} 

\author{Niko~Jokela}
\email{niko.jokela@helsinki.fi}
\affiliation{Department of Physics,
P.O.~Box 64, FI-00014 University of Helsinki, Finland}
\affiliation{Helsinki Institute of Physics,
P.O.~Box 64, FI-00014 University of Helsinki, Finland}

\author{K.~Kajantie}
\email{keijo.kajantie@helsinki.fi}
\affiliation{Department of Physics,
P.O.~Box 64, FI-00014 University of Helsinki, Finland}
\affiliation{Helsinki Institute of Physics,
P.O.~Box 64, FI-00014 University of Helsinki, Finland}

\author{M.~Laine}
\email{laine@itp.unibe.ch}
\affiliation{AEC, 
ITP, 
University of Bern, 
Sidlerstrasse 5, CH-3012 Bern, Switzerland}

\author{Sami~Nurmi}
\email{sami.t.nurmi@jyu.fi}
\affiliation{Department of Physics,
P.O.~Box 35, FI-40014 University of Jyv\"askyl\"a, Finland}
\affiliation{Helsinki Institute of Physics,
P.O.~Box 64, FI-00014 University of Helsinki, Finland}

\author{Miika~Sarkkinen}
\email{miika.sarkkinen@helsinki.fi}
\affiliation{Department of Physics,
P.O.~Box 64, FI-00014 University of Helsinki, Finland}
\affiliation{Helsinki Institute of Physics,
P.O.~Box 64, FI-00014 University of Helsinki, Finland}

\begin{abstract}
The free propagator of a massless mode in an expanding universe can be 
written as a sum of two terms, a lightcone and a tail part.
The latter describes a subluminal (time-like) signal. 
We show that the 
inflationary gravitational wave background, influencing 
cosmic microwave background polarization, and routinely used 
for constraining inflationary models through the so-called $r$ ratio, 
originates exclusively from the tail part. 
\end{abstract}

\date{October 2023}

\preprint{HIP-2023-8/TH}


\maketitle


\section{Introduction}
\la{se:intro}

In four-dimensional flat spacetime, massless radiation propagates
along light cones. This means that the corresponding retarded Green's
function is proportional to the Dirac $\delta$-function. On the other
hand, in curved spacetime, massless fields develop a so-called tail. 
This means that the Green's function is non-zero inside the lightcone. 

The appearance of a tail was first observed in a mathematical analysis
of second order partial differential equations~\cite{hadamard,friedlander}.
Subsequently it has been studied in a multitude of physical contexts~\cite{%
kk3,kk4,kk5,kk6,kk7,kk8,kk9,kk10,kk11,kk12,kk13,kk14,kk15,%
kk16,kk17,kk18,kk19,kk20,kk21,kk22,kk23,kk24,kk25}, 
suggesting possible observable signatures. However, so far there is 
no direct observation of the tail. 

A particularly relevant curved spacetime is the cosmological 
Friedmann-Lema\^itre-Robertson-Walker (FLRW) universe. A fundamental
observation from its late period is the gravitational 
wave (GW) signal from mergers of black holes~\cite{ligo}. The main signal 
arrives on the lightcone and 
is theoretically well understood~\cite{ligo_gr}.
The corresponding tail has not yet been observed, but its
magnitude has been computed~\cite{kk22}
(see also refs.~\cite{kk17,kk18,kk19,kk20,kk21}).
The tail signal arrives long after the main merger signal, 
but its magnitude is surprisingly large. 

The FLRW universe also contains earlier periods. A particularly
important one is that of inflation, where the seeds for structure formation, 
as well as the anisotropies that are observed in the cosmic microwave
background (CMB)~\cite{planck}, are believed to have been generated. 
At linear order 
these originate from so-called scalar perturbations, but neither the scalar
field driving inflation nor its perturbations are massless. 
However, inflation also generates
tensor perturbations~\cite{gw1,gw2,gw3,gw4}, which manifest themselves
as gravitational waves. 
The gravitational wave background propagates
until today, in a well-understood fashion~\cite{sw}, and leads 
to potentially observable consequences, through the 
polarization of the CMB photons~\cite{sz}. In principle primordial GWs 
could also be observed directly, e.g.\ via pulsar timing
arrays~\cite{nhz1,nhz2,nhz3,nhz4}, 
even if the sensitivity is not sufficient for the 
simplest inflationary models~\cite{sv}. 

As the spacetime curvature is large during inflation, 
we may also anticipate a tail contribution from this epoch. 
We now proceed to describing
the computation of the inflationary GW background, quantifying
subsequently the tail's role in it. 

%
\section{Gravitational wave background from de Sitter vacuum fluctuations}
\la{sec:stochastic}

The way that the inflationary GW background is computed
is that we first consider the wave equation for tensor perturbations
in de Sitter spacetime. The time dependence of the solution is easily
found, but its normalization needs also to be fixed.
This can be done by considering a fixed co-moving momentum, $k$. 
Thanks to the expansion of the universe, the corresponding physical
wavelength is very small at early times. Therefore, it is within 
a causally connected domain, ``inside the horizon''. 
There its normalization can be fixed like in a Minkowskian vacuum, 
a well-established problem in quantum field theory. 

Once the normalization and time evolution have been found, we can 
follow the momenta until late times. At some point, the modes 
``exit the horizon'', i.e.\ $k \ll a H$ (or, in terms
of physical momenta $p \equiv k/a$, $p \ll H$). Here $H$ is the Hubble
rate, which is constant in de Sitter spacetime, and $a$ is the 
scale factor, which grows exponentially in physical time. 

Once the modes exit the horizon, they ``freeze out'', i.e.\ their
amplitude becomes constant. Up to overall normalization, 
the absolute value squared of the amplitude constitutes the
primordial tensor power spectrum 
that we are interested in ($\P^{ }_\T$). 

The modes do not stay forever outside of the horizon. As inflation ends, 
the Hubble constant starts to decrease, and the scale factor grows less
rapidly. At some point, a given momentum mode re-enters the horizon. 
Then it starts to oscillate again. As a result, it carries energy
density, which is in principle observable.
It also influences other modes propagating through the cosmological
history, notably CMB photons. 
The determination of these post-inflationary features amounts 
to the determination of a ``transfer function'' 
from the primordial to the current era~\cite{sw}.
We will not concern ourselves with the complicated 
post-inflationary physics, 
only the primordial power spectrum. 

\begin{figure}[t]

\hspace*{-0.1cm}
\centerline{%
   ~~~\epsfxsize=7.5cm\epsfbox{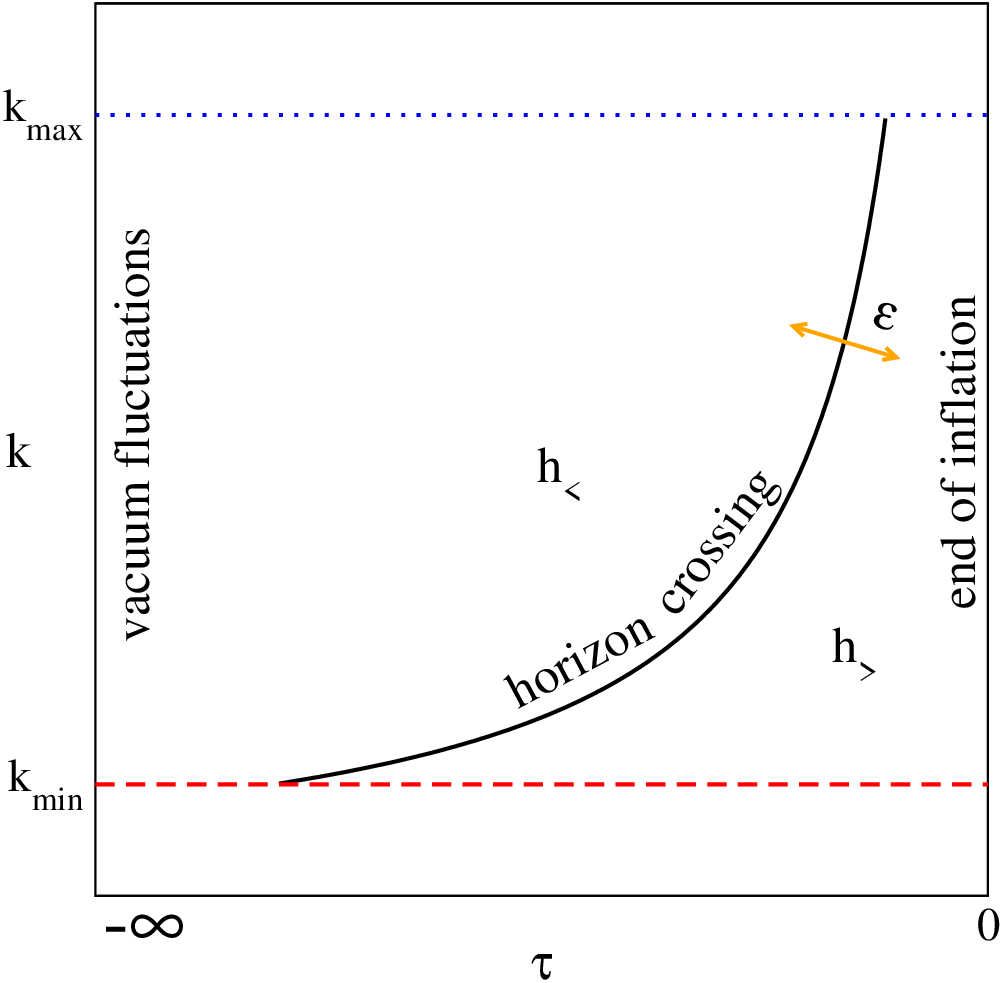}
}

\caption[a]{\small
  An illustration of the computation of the gravitational wave background
  with the stochastic formalism, in the plane of conformal time ($\tau$) and
  comoving momentum ($k$). The quantum mechanical solution at early times
  is denoted by~$h^{ }_<$. A chosen ``horizon crossing'' 
  hypersurface (whose position $k = -\epsilon/\tau$ depends 
  on a parameter $\epsilon$ that cancels from final results)
  is used for matching $h^{ }_<$ onto a long-wavelength field 
  $h^{ }_>$. The latter can be treated as a classical perturbation. 
}

\la{fig:horizon}
\end{figure}

In order to compute $\P^{ }_\T$, we find it illuminating to 
carry out the computation with the formalism of stochastic 
inflation~\cite{stochastic}. As illustrated in \fig\ref{fig:horizon}, 
the full time evolution is then divided into two parts. While the 
early part is effectively the same as in the standard 
quantum-mechanical analysis, the latter part is simpler, as
it can be viewed as a classical problem. This facilitates the 
identification of the lightcone and tail contributions 
to the final result. 

The stochastic formalism is employed 
in different variants in the literature.
In the following, we implement it in a way that is not an 
approximation but just a mathematical re-organization of the 
usual quantum-mechanical computation (cf., e.g., refs.~\cite{yn,rr,scan}
and references therein). Apart from a clear physical picture, 
the advantage of this approach is that it introduces an arbitrary
parameter, denoted by~$\epsilon$, whose cancellation 
offers for a nice crosscheck,
analogously to the role played by 
the gauge parameter in Yang-Mills theories. 

Let $h$ be a canonically normalized massless field. 
Denoting by $\tau \in (-\infty,0)$ the conformal time, de Sitter
spacetime with a constant Hubble parameter $H$ has the scale factor
$
 a = -1 / (H \tau)
$.
The solution of the wave equation
$
  h''_{ } - \frac{2}{\tau} h'_{ }
 - \nabla^2 h \; = \; 0
$
reads 
\be
 h^{ }_{ } 
 \; = \; 
 \int \! \frac{{\rm d}^3\vec{k}}{\sqrt{(2\pi)^{3}_{ }}} 
 \Bigl[ \,
    w^{ }_\rmii{\vec{k}}
    \hspace*{-8mm} 
 \underbrace{ h^{ }_k(\tau) }_{
   \frac{ i H }{ \sqrt{2 k^3} }
   (1 + i k \tau)\, e^{-i k \tau}
 } 
    \hspace*{-8mm} 
    e^{ i \vec{k}\cdot\vec{x}}_{}   
  + 
    \hc
 \, \Bigr]
 \;, \la{scalar}
\ee
where $\vec{k}$ and $\vec{x}$ are a comoving momentum and coordinate, 
respectively, 
$
    w^{ }_\rmii{\vec{k}}
$ is an annihilation operator of the distant-past (Bunch-Davies) vacuum, 
and the canonical commutation relation takes the form
$
  [\, w^{ }_\rmii{\vec{k}} , w^{\dagger}_\rmii{\vec{l}} \,]
  \; \equiv \; 
  \delta^{(3)}(\vec{k-l})
$.

The mode $h$ is now divided into short-distance
($h^{ }_{<}$) and long-distance parts ($h^{ }_{>}$), 
$
 h =  h^{ }_{>} + h^{ }_{<}
$, 
by defining
\be
 h^{ }_{<} 
 \;\equiv\; 
 \int \! \frac{{\rm d}^3\vec{k}}{\sqrt{(2\pi)^{3}_{ }}} 
 \;\; 
 \underbrace{ W^{ }_k(\tau) }_{
   \theta(k + \frac{\epsilon}{\tau})
 } \;\; 
 \Bigl[ \,
    w^{ }_\rmii{\vec{k}}
    \, h^{ }_k(\tau) 
    \, e^{ i \vec{k}\cdot\vec{x}}_{}   
  + 
    \hc
 \, \Bigr]
 \;, \la{splitup}
\ee
where the window function $W^{ }_k$ selects large momenta. 
The parameter 
$\epsilon$ is arbitrary and must 
drop out from physical results~\cite{footnote}.
Inserting \eq\nr{splitup} into the equation of motion yields
\ba
 h''_{>} - \frac{2}{\tau}\, h'_{>}
 - \nabla^2 h^{ }_{>} & = & \xiQ^{ }
 \;, \la{eom_h>} \\
 \xiQ^{ } & \equiv & 
 - \biggl( 
       \partial_\tau^2 - \frac{2}{\tau}\, \partial^{ }_\tau - \nabla^2
   \biggr)
  \, h^{ }_{<}
 \;, \la{noise}
\ea
where the ``quantum noise'' has the form 
\be
 \xiQ^{ }(\tau,\vec{x})  =  
 - \int \! \frac{{\rm d}^3\vec{k}}{\sqrt{(2\pi)^{3}_{ }}} \, 
 \Bigl[ \,
    w^{ }_\rmii{\vec{k}}
    \hspace*{-1.0cm}
    \underbrace{ f^{ }_k(\tau) }_
    {
    \bigl( W''_{k} - \frac{2}{\tau}\, W'_{k} \bigr)
    h^{ }_k 
   + 2 W'_{k} h'_k
    }
    \hspace*{-1.0cm}
    e^{ i \vec{k}\cdot\vec{x}}_{}   
  + 
    \hc
 \, \Bigr]
 \;. \la{rho_Q} 
\ee

We note that with $W^{ }_k$ from \eq\nr{splitup}, the 
function $f^{ }_k$ in \eq\nr{rho_Q} contains the Dirac-$\delta$ or 
its derivative. Therefore, the differential operator acting
on $h^{ }_{<}$ amounts to a holographic projection, localizing
the information from early times onto the horizon crossing
hypersurface (cf.\ \fig\ref{fig:horizon}). 

The long-wavelength mode $h^{ }_{>}$ can subsequently be determined 
from \eq\nr{eom_h>} with a retarded Green's function. The Green's 
function satisfies
\be
 \biggl(\, 
  \partial_\tau^2 - \frac{2}{\tau}\, \partial^{ }_\tau -\nabla^2_\vec{x} 
 \,\biggr) \, 
 G^{ }_{|\vec{x-z}|}(\tau,\taui^{ }) = 
 \delta(\tau - \taui^{ })\,\delta^{(3)}_{ }(\vec{x-z}) 
 \;, \la{greens1}
\ee
with the boundary conditions 
\ba
 G^{ }_{|\vec{x-z}|}(\tau,\taui^{ })
 & \stackrel{ \tau \le \taui^{ } }{=}  & 
 0 
 \;, \\
 \lim^{ }_{\tau\to\taui^{+}}
 \partial^{ }_\tau 
 G^{ }_{|\vec{x-z}|}(\tau,\taui^{ }) 
 & = &
 \delta^{(3)}_{ }(\vec{x-z})
 \;. \la{greens2}
\ea
Here 
$\tau$ and $\taui^{ }$ are the time
arguments of observation and source, respectively.

Given the Green's function, 
we can determine the late-time solution as
\be
 h^{ }_{>}(\tau,\vec{x})
 = 
 \int^{ }_\vec{z}
 \int_{-\infty}^{\tau} \! {\rm d}\tau^{ }_1 \, 
 G^{ }_{|\vec{x-z}|}(\tau,\tau^{ }_1)\,
 \xiQ^{ }(\tau^{ }_1,\vec{z})
 \;, \la{h_>}
\ee
where $\int^{ }_\vec{z}$ is a spatial integral over source locations.
This can be viewed as an anti-holographic mapping, from the hypersurface
unto late times. Our next task is then to specify the properties
of the Green's function. 

%
\section{Lightcone and tail parts of the Green's function}
\la{sec:greens}

The Green's function from 
\eqs\nr{greens1}--\nr{greens2} is easily solved by representing it 
in momentum space,
\be
 G^{ }_x(\tau,\tau^{ }_i) = 
 \int_\vec{k} e^{i\vec{k}\cdot\vec{x}}_{ }
 \, 
 G^{ }_k(\tau,\tau^{ }_i)
 \;,
\ee 
where $\int_\vec{k} \equiv \int\!\frac{{\rm d}^3\vec{k}}{(2\pi)^3}$. 
The momentum-space solution is 
split up into two parts, called the lightcone (lc) 
and the tail parts (tail), defined as  
\ba
 G^{ }_k(\tau,\tau^{ }_i)
 & = & 
 \theta(\tau - \tau^{ }_i)\, 
 \bigl[\, 
   g^{ }_{k,\rmi{lc}}(\tau,\tau^{ }_i) + 
   g^{ }_{k,\rmi{tail}}(\tau,\tau^{ }_i)
 \,\bigr]
 \;, \la{G_k} \hspace*{4mm} \\[2mm] 
 g^{ }_{k,\rmi{lc}}(\tau,\tau^{ }_i) & \equiv & 
 \frac{\tau \, \sin [k(\tau - \tau^{ }_i)]}{ k \tau^{ }_i} 
 \;, \la{g_k_lc} \\ 
 g^{ }_{k,\rmi{tail}}(\tau,\tau^{ }_i) & \equiv & 
 \frac{
    \sin [k(\tau - \tau^{ }_i)] 
    - k (\tau - \tau^{ }_i)
    \cos [k(\tau - \tau^{ }_i)]
 }{k^3 \tau_i^2}
 \;. \nn \la{g_k_tail}
\ea
The rationale behind this nomenclature becomes clear when we go  
back to configuration space.
Then we obtain
\ba
 g^{ }_{x,\rmi{lc}}(\tau,\tau^{ }_i)
 \quad & 
 \underset{ x>0 }{  
 \overset{ \tau > \tau^{ }_i }{ = } } 
 & \quad
 \frac{\tau\,\delta(\tau-\tau^{ }_i-x)}{4\pi x \tau^{ }_i}
 \;, \la{g_x_lc} \\
 g^{ }_{x,\rmi{tail}}(\tau,\tau^{ }_i)
 \quad &
 \underset{ x>0 }{  
 \overset{ \tau > \tau^{ }_i }{ = } } 
 & \quad
 \frac{\theta(\tau-\tau^{ }_i-x)}{4\pi \tau^{2}_i}
 \;. \la{g_x_tail}
\ea
The lightcone part describes a signal arriving at the speed of 
light, the tail part arrives later. 
It can be verified that the sum of the two parts,
though not the parts separately, 
satisfies the $s$-wave equation
\be
 \biggl(
   \partial_\tau^2 - \frac{2}{\tau} \partial^{ }_\tau 
 - \partial_x^2 - \frac{2}{x} \partial^{ }_x 
 \biggr)
 \bigl(
  g^{ }_{x,\rmi{lc}} + g^{ }_{x,\rmi{tail}}
 \bigr)
  = 0  
 \;. 
\ee
The $s$-wave solution is the relevant one, because the source 
in \eq\nr{greens1} is a monopole. 

%
\section{Computation of the primordial tensor power spectrum}
\la{sec:k-space}

In cosmology, the fundamental objects are equal-time 
2-point correlation functions, such as those of the temperature
of the CMB photons. 
What we are interested in here is the 2-point correlator
of the tensor perturbations. For the $h^{ }_{>}$-field, 
from \eq\nr{h_>}, evaluating the expectation value
in the distant-past vacuum, 
the equal-time correlator can be expressed as 
\ba
 && \hspace*{-0.6cm}
 \langle\, 
 h^{ }_{>}(\tau,\vec{x}) \, 
 h^{ }_{>}(\tau,\vec{y}) 
 \,\rangle
 =  \nn[2mm]
 &&
 \int_{\vec{z,w}}
 \int_{-\infty}^{\tau} \!\!\! {\rm d}\tau^{ }_1 \, 
 \int_{-\infty}^{\tau} \!\!\! {\rm d}\tau^{ }_2 \, 
 \, 
 G^{ }_{|\vec{x-z}|}(\tau,\tau^{ }_1)\,
 G^{ }_{|\vec{y-v}|}(\tau,\tau^{ }_2)
 \,
 \nn[2mm] &&
 \, \times \, 
 \langle 0 | \, 
 \xiQ^{ }(\tau^{ }_1,\vec{z}) \, 
 \xiQ^{ }(\tau^{ }_2,\vec{v}) 
 \, | 0 \rangle 
 \;. \la{hx_hy} 
\ea
The quantum mechanics of the problem is now hidden in the 
autocorrelator of~$\xiQ^{ }$. 
Inserting \eq\nr{rho_Q}, it becomes 
\be
 \langle 0 | \, 
 \xiQ^{ }(\tau^{ }_1,\vec{z}) \, 
 \xiQ^{ }(\tau^{ }_2,\vec{v}) 
 \, | 0 \rangle 
 \; = \; 
 \int_\vec{k}
 e^{i\vec{k}\cdot(\vec{z} - \vec{v})} 
 \, 
 f^{ }_k (\tau^{ }_1) \, f^*_k (\tau^{ }_2) 
 \;. \la{noise_qm}
\ee

The correlator of \eq\nr{hx_hy}
is most easily evaluated in co-moving momentum space
(the configuration space computation is described
in the supplementary material). 
After a Fourier transform, we obtain
\ba
 && \hspace*{-1.0cm}
 \langle\, 
 h^{ }_{>}(\tau,\vec{k}) \, 
 h^{ }_{>}(\tau,\vec{q}) 
 \,\rangle
 \nn[2mm]
 & = & \;
 \deltabar(\vec{k+q})
 \, 
 \biggl| 
  \int_{-\infty}^{\tau} \, {\rm d}\taui^{ } \, 
  G^{ }_k(\tau,\taui^{ }) \, f^{ }_k(\taui^{ })
 \biggr|^2_{ }
 \;, \la{hk_hq}
\ea
where $\int_\vec{k}\,\deltabar(\vec{k}) \equiv 1$.

Making use of $f^{ }_k$ from \eq\nr{rho_Q}, 
and carrying out a partial integration, 
the integral in \eq\nr{hk_hq} can be
expressed as 
\ba
 && \hspace*{-1.3cm} 
  \int_{-\infty}^{\tau} 
  \, {\rm d}\taui^{ } \, 
  G^{ }_k(\tau,\taui^{ }) 
 \, f^{ }_k(\taui^{ })
 \nn 
 & = & 
 \int_{-\infty}^{\tau} \, {\rm d}\taui^{ } \, W'_k(\taui^{ }) \, 
 \biggl\{ 
   - \partial^{ }_{\taui^{ }} G^{ }_k(\tau,\taui^{ })
     h^{ }_k(\taui^{ })
 \nn 
 & & \,  
   + \; G^{ }_k(\tau,\taui^{ })
   \biggl[
    h'_k(\taui^{ }) -  
    \frac{2}{\taui^{ }}\, h^{ }_k(\taui^{ }) 
   \biggr]
 \biggr\}
 \;. \la{G_R_full} 
\ea
With 
$ 
 W_k' = -\frac{\epsilon}{\tau^2}\,
 \delta\bigl( k + \frac{\epsilon}{\tau}\bigr) 
$, the lightcone part yields
\ba
 &&
  \int_{-\infty}^{\tau} 
  \, {\rm d}\taui^{ } \, 
  g^{ }_{k,\rmi{lc}}(\tau,\taui^{ }) 
 \, f^{ }_k(\taui^{ })
 = 
 \frac{i H k\tau}{\sqrt{2k^3}} 
 \frac{e^{i \epsilon}_{}}{\epsilon}
  \la{g_f_lc} \\
 && \, \times \; 
 \biggl[
   \cos(k\tau + \epsilon)\, \bigl( 1 - i \epsilon \bigr)
 + \sin(k\tau + \epsilon)\, \biggl( \frac{1}{\epsilon} - i -\epsilon \biggr)
 \biggr] 
 \;, \nonumber
\ea
whereas for the tail part we obtain
\ba
 &&
  \int_{-\infty}^{\tau} 
  \, {\rm d}\taui^{ } \, 
  g^{ }_{k,\rmi{tail}}(\tau,\taui^{ }) 
 \, f^{ }_k(\taui^{ })
 = 
 \frac{i H k\tau}{\sqrt{2k^3}} 
 \frac{e^{i \epsilon}_{}}{\epsilon}
 \la{g_f_tail} \\ 
 && \times \! 
 \biggl[
   \cos(k\tau + \epsilon) \biggl(\! -1 - \frac{\epsilon}{k\tau} \!\biggr)\!
  + \sin(k\tau + \epsilon) \biggl(\! -\frac{1}{\epsilon} + i
  + \frac{i\epsilon}{k\tau} \biggr)
 \biggr] 
 \,. \nonumber
\ea
Summing together, many terms cancel, 
and the remaining ones can be factorized, 
\ba
 && 
  \int_{-\infty}^{\tau} 
  \, {\rm d}\taui^{ } \, 
  G^{ }_{k}(\tau,\taui^{ }) 
 \, f^{ }_k(\taui^{ })
 \nn 
 & = &  
 \frac{i H k\tau}{\sqrt{2k^3}} 
 \bigl( - e^{i \epsilon}_{} \bigr)
 \biggl( i + \frac{1}{ k \tau }\biggr)
 \bigl[
   \cos(k\tau + \epsilon)
  - i \sin(k\tau + \epsilon)
 \bigr] 
 \nn 
 & = & 
 -\frac{i H}{\sqrt{2k^3}} 
 e^{-i k \tau}_{}
 \bigl( 1 + i k \tau\bigr)
 \;. \la{g_f_sum}
\ea
In the equations above, an implicit factor $\theta(1/\tau + k/\epsilon)$
has been suppressed for simplicity of notation. 

An important property of \eq\nr{g_f_sum} is that it is independent
of the parameter $\epsilon$, and thus of the position of the 
horizon crossing hypersurface in \fig\ref{fig:horizon}. That is, 
the anti-holographic mapping from the hypersurface onto late 
times erases all memory of the hypersurface itself. 

We note that \eq\nr{g_f_sum} is 
just the standard result for massless mode functions. The novelty of
our computation is that the first term in the parentheses, 
dominant on superhorizon scales $k|\tau| \ll 1$, 
is seen to come exclusively
from the tail, the second from the lightcone. 

For physical conclusions, we need 
the absolute value squared of \eq\nr{g_f_sum}, 
according to \eq\nr{hk_hq}.
The corresponding power spectrum is obtained by  
multiplying this with the measure $k^3/(2\pi^2)$, yielding
\be 
 \P^{ }_{h}
 \; = \; 
 \biggl( \frac{H}{2\pi} \biggr)^2_{ }
 \bigl(\, 1 + k^2\tau^2 \,\bigr)
 \;. \la{P_h}
\ee
After adding the normalization factors
for tensor perturbations, 
and considering the limit $k|\tau| = k/(aH) \ll 1$, valid
for momenta well outside of the horizon (these are the ones having
an observable effect today),  
this reduces to the text-book tensor power spectrum, 
\be
 \P^{ }_\rmii{T}
 \; \stackrel{k|\tau| \ll 1}{\approx} \; 
 \frac{16}{\pi}
 \biggl( \frac{H}{\mpl^{ }} \biggr)^2_{ }
 \;, \la{P_T}
\ee
where $\mpl^{ } \equiv 1.22091 \times 10^{19}_{ }$~GeV
is the Planck mass. 

{}From the tensor power spectrum, once multiplied
with the transfer function~\cite{sw}, we can 
obtain the current-day fractional 
gravitational energy density, 
$\Omega^{ }_\rmii{GW}$. 
Measuring this is an on-going effort~\cite{nhz1,nhz2,nhz3,nhz4}. 
The ratio of the tensor power spectrum to the curvature one, 
$
 r \equiv \P^{ }_\rmii{T} / \P^{ }_\rmii{$\mathcal{R}$}
$, 
is already strongly 
constrained by Planck data on CMB, $r < 0.056$~\cite{planck}.

We stress that the only terms left over in \eq\nr{P_T}, 
are the ones enhanced by $1/(k\tau)$ in \eq\nr{g_f_tail}. 
In other words, \eq\nr{P_T}
arises exclusively from the tail contribution. However, the lightcone
contribution is also conceptually important, 
for it guarantees the independence of the result of 
the parameter~$\epsilon$ at finite values 
of $k\tau$~\cite{remark}
(cf.\ also \eqs\nr{x_penul} and \nr{x_final} 
in the supplementary material).

%
\section{Conclusions}

The purpose of this paper has been to demonstrate that 
the primordial tensor power spectrum, cf.\ \eq\nr{P_T}, which
plays an important role in constraining inflationary models 
through CMB data~\cite{planck}, originates from the
$1/(k\tau)$-enhanced terms in \eq\nr{g_f_tail}. In other words, 
the tails of the gravitational waves that cross the horizon
are responsible for the physical phenomena observable today, 
implementing thereby a remarkable memory effect.

%
\section*{Acknowledgements}

N.J.\ and M.S.\ have been supported in part by the Academy
of Finland, grant no.\ 1322307, 
and M.L.\ by the Swiss National Science Foundation
(SNSF), under grant 200020B-188712.

%

\bibliographystyle{apsrev4-1}

\begin{thebibliography}{99}

\bibitem{hadamard}
  J.~Hadamard, 
  {\it Lectures on Cauchy's Problem in Linear Partial Differential Equations}
  (Yale University Press, New Haven, 1923). 

\bibitem{friedlander}
  F.G.~Friedlander, 
  {\it The Wave Equation on a Curved Space-Time}
  (Cambridge University Press, Cambridge, 2010). 

\bibitem{kk3}
  B.S.~DeWitt and R.W.~Brehme,
  {\it Radiation damping in a gravitational field,}
  Annals Phys.\ {9} (1960) 220.

\bibitem{kk4}
  R.H.~Price,
  {\it Nonspherical Perturbations of Relativistic Gravitational Collapse.
  I. Scalar and Gravitational Perturbations,}
  Phys.\ Rev.\ D {5} (1972) 2419.

\bibitem{kk5}
  L.~Blanchet and T.~Damour,
  {\it Tail-transported temporal correlations in the dynamics
  of a gravitating system,}
  Phys.\ Rev.\ D {37} (1988) 1410.

\bibitem{kk15}
  T.W.~Noonan,
  {\it Huygens's Principle for the Electromagnetic Vector 
  Potential in Riemannian Spacetimes,}
  Astrophys.\ J.\ {341} (1989) 786.

\bibitem{kk7}
  V.~Faraoni and S.~Sonego,
  {\it On the tail problem in cosmology,}
  Phys.\ Lett.\ A {170} (1992) 413
  [astro-ph/9209004].

\bibitem{kk6}
  A.G.~Wiseman,
  {\it Coalescing binary systems of compact objects to 
  $\mbox{(post)}^{\fr52}_{ }$-Newtonian order. IV. 
  The gravitational wave tail,}
  Phys.\ Rev.\ D {48} (1993) 4757

\bibitem{kk8}
  R.R.~Caldwell,
  {\it Green's functions for gravitational waves in FRW spacetimes,}
  Phys.\ Rev.\ D {48} (1993) 4688
  [gr-qc/9309025].

\bibitem{kk23}
  L.~Bombelli and S.~Sonego,
  {\it Relationships between various characterizations of wave tails,}
  J.\ Phys.\ A {27} (1994) 7177
  [math-ph/0002026].

\bibitem{kk9}
  J.~Iliopoulos, T.N.~Tomaras, N.C.~Tsamis and R.P.~Woodard,
  {\it Perturbative quantum gravity and Newton's law
  on a flat Robertson-Walker background,}
  Nucl.\ Phys.\ B {534} (1998) 419
  [gr-qc/9801028].

\bibitem{kk10}
  H.J.~de Vega, J.~Ramirez and N.G.~Sanchez,
  {\it Generation of gravitational waves by generic sources
  in de Sitter space-time,}
  Phys.\ Rev.\ D {60} (1999) 044007
  [astro-ph/9812465].

\bibitem{kk11}
  V.~Faraoni and E.~Gunzig,
  {\it Tales of tails in cosmology,}
  Int.\ J.\ Mod.\ Phys.\ D {08} (1999) 177
  [astro-ph/9902262].

\bibitem{kk12}
  V.~Balek and V.~Polak,
  {\it Group velocity of gravitational waves in an expanding universe,}
  Gen.\ Rel.\ Grav.\ {41} (2009) 505
  [0707.1513].

\bibitem{kk14}
  A.I.~Harte,
  {\it Tails of plane wave spacetimes:
  Wave-wave scattering in general relativity,}
  Phys.\ Rev.\ D {88} (2013) 084059
  [1309.5020].

\bibitem{kk24}
  A.~Blasco, L.J.~Garay, M.~Martin-Benito and E.~Martin-Martinez,
  {\it Violation of the strong Huygen's principle and timelike signals
  from the early universe,}
  Phys.\ Rev.\ Lett.\ {114} (2015) 141103
  [1501.01650].

\bibitem{kk25}
  A.~Blasco, L.J.~Garay, M.~Martin-Benito and E.~Martin-Martinez,
  {\it Timelike information broadcasting in cosmology,}
  Phys.\ Rev.\ D {93} (2016) 024055
  [1510.04701].

\bibitem{kk13}
  A.~Ashtekar, B.~Bonga and A.~Kesavan,
  {\it Asymptotics with a positive cosmological constant. III.
  The quadrupole formula,}
  Phys.\ Rev.\ D {92} (2015) 104032
  [1510.05593].

\bibitem{kk18}
  A.~Kehagias and A.~Riotto,
  {\it BMS in cosmology,}
  JCAP {05} (2016) 059
  [1602.02653].

\bibitem{kk20}
  Y.-Z.~Chu,
  {\it Gravitational wave memory in dS$_{4+2n}$ and 4D cosmology,}
  Class.\ Quant.\ Grav.\ {34} (2017) 035009
  [1603.00151].

\bibitem{kk17}
  A.~Tolish and R.M.~Wald,
  {\it Cosmological memory effect,}
  Phys.\ Rev.\ D {94} (2016) 044009
  [1606.04894].

\bibitem{kk19}
  Y.-Z.~Chu,
  {\it More on cosmological gravitational waves and their memories,}
  Class.\ Quant.\ Grav.\ {34} (2017) 194001
  [1611.00018].

\bibitem{kk21}
  M.A.~Ismail, Y.-Z.~Chu and Y.-W.~Liu,
  {\it Late time tails and nonlinear memories in 
  asymptotically de Sitter spacetimes,}
  Phys.\ Rev.\ D {104} (2021) 104038
  [2101.01736].

\bibitem{kk22}
  N.~Jokela, K.~Kajantie and M.~Sarkkinen,
  {\it Gravitational wave memory and its tail in cosmology,}
  Phys.\ Rev.\ D {106} (2022) 064022
  [2204.06981].

\bibitem{kk16}
  J.S.~Santos, V.~Cardoso and J.~Nat\'ario,
  {\it Electromagnetic radiation reaction and energy extraction
  from black holes: The tail term cannot be ignored,}
  Phys.\ Rev.\ D {107} (2023) 064046
  [2303.03411].

\bibitem{ligo}
  B.P.~Abbott {\it et al.} [LIGO Scientific and Virgo Collaborations],
  {\it Observation of gravitational waves from a binary black hole merger,}
  Phys.\ Rev.\ Lett.\ {116} (2016) 061102
  [1602.03837].

\bibitem{ligo_gr}
  B.P.~Abbott {\it et al.} [LIGO Scientific and Virgo Collaborations],
  {\it Tests of general relativity with GW150914,}
  Phys.\ Rev.\ Lett.\ {116} (2016) 221101; 
  {\it ibid.}\ {121} (2018) 129902 (E)
  [1602.03841].

\bibitem{planck}
  Y.~Akrami {\it et al} [Planck Collaboration],
  {\it Planck 2018 results.\ X.\ Constraints on inflation},
  Astron.\ Astrophys.\ {641} (2020) A10
  [1807.06211].

\bibitem{gw1}
  L.P.~Grishchuk,
  {\it Amplification of gravitational waves in an isotropic universe,}
  Sov.\ Phys.\ JETP 40 (1975) 409
  [Zh.\ Eksp.\ Teor.\ Fiz.\ {67} (1974) 825].

\bibitem{gw2} 
  A.A.~Starobinsky,
  {\it Spectrum of relict gravitational radiation and
  the early state of the universe,}
  JETP Lett.\  {30} (1979) 682
  [Pisma Zh.\ Eksp.\ Teor.\ Fiz.\  {30} (1979) 719].

\bibitem{gw3}
  V.A.~Rubakov, M.V.~Sazhin and A.V.~Veryaskin,
  {\it Graviton creation in the inflationary universe
  and the grand unification scale,}
  Phys.\ Lett.\ B {115} (1982) 189.

\bibitem{gw4}
  R.~Fabbri and M.D.~Pollock,
  {\it The effect of primordially produced gravitons upon
  the anisotropy of the cosmological microwave background radiation,}
  Phys.\ Lett.\ B {125} (1983) 445.

\bibitem{sw}
  S.~Weinberg,
  {\it Damping of tensor modes in cosmology,}
  Phys.\ Rev.\ D {69} (2004) 023503
  [astro-ph/0306304].

\bibitem{sz}
  U.~Seljak and M.~Zaldarriaga,
  {\it Signature of gravity waves in polarization
  of the microwave background,}
  Phys.\ Rev.\ Lett.\ {78} (1997) 2054
  [astro-ph/9609169].

\bibitem{nhz1}
  G.~Agazie {\it et al.},
  {\it The NANOGrav 15 yr Data Set: Evidence for a 
  gravitational-wave background,}
  Astrophys.\ J.\ Lett.\ {951} (2023) L8
  [2306.16213].

\bibitem{nhz2}
  J.~Antoniadis {\it et al.}, 
  {\it The second data release from the European Pulsar Timing Array III.
  Search for gravitational wave signals,}
  Astron.\ Astrophys.\ {678} (2023) A50
  [2306.16214].

\bibitem{nhz3}
  D.J.~Reardon {\it et al.},
  {\it Search for an isotropic gravitational wave background
  with the Parkes Pulsar Timing Array,}
  Astrophys.\ J.\ Lett.\ {951} (2023) L6
  [2306.16215].

\bibitem{nhz4}
  H.~Xu {\it et al.}, 
  {\it Searching for the nano-Hertz stochastic gravitational wave background
  with the Chinese Pulsar Timing Array data release I,}
  Res.\ Astron.\ Astrophys.\ {23} (2023) 075024
  [2306.16216].

\bibitem{sv}
  S.~Vagnozzi,
  {\it Implications of the NANOGrav results for inflation,}
  Mon.\ Not.\ Roy.\ Astron.\ Soc.\ {502} (2021) L11
  [2009.13432].

\bibitem{stochastic}
  A.A.~Starobinsky and J.~Yokoyama,
  {\it Equilibrium state of a self-interacting scalar field
  in the de Sitter background,}
  Phys.\ Rev.\ D {50} (1994) 6357
  [astro-ph/9407016].

\bibitem{yn}
  M.~Sasaki, Y.~Nambu and K.-i.~Nakao,
  {\it Classical behavior of a scalar field in the inflationary universe,}
  Nucl.\ Phys.\ B {308} (1988) 868.

\bibitem{rr}
  R.O.~Ramos and L.A.~da Silva,
  {\it Power spectrum for inflation models with quantum and thermal noises,}
  JCAP {03} (2013) 032
  [1302.3544].

\bibitem{scan}
  P.~Klose, M.~Laine and S.~Procacci,
  {\it Gravitational wave background from vacuum and
  thermal fluctuations during axion-like inflation,}
  JCAP {12} (2022) 020
  [2210.11710].

\bibitem{footnote}
  As alluded to above, in the literature
  the stochastic formalism is often implemented
  in an approximate fashion, in which case it is only correct for 
  $\epsilon \ll 1$.

\bibitem{remark}
  We note that if we expand \eq\nr{g_f_sum} in 
  a small $k\tau$, then the leading term is the constant, 
  and the next-to-leading term is of $\rmO(k^2\tau^2)$. 
  If we take a time derivative, as is relevant for energy
  density, the leading term drops out. Thus, the tail 
  part does {\em not} dominate the would-be energy density
  measured outside of the horizon (this is a purely 
  theoretical construct). But it is the constant part
  that turns into the energy density observable today~\cite{sw}. 

\end{thebibliography}

%

\onecolumngrid
\appendix
\newpage

%
\section{Supplementary material: Evaluation in position space}
\la{sec:x-space}

In the body of the text we have considered the 2-point correlator
of tensor perturbations in momentum space, cf.\ \eq\nr{hk_hq}. 
The original position-space correlator of \eq\nr{hx_hy} can then 
be obtained from a Fourier transform, 
\ba
 \langle\, 
 h^{ }_{>}(\tau,\vec{x}) \, 
 h^{ }_{>}(\tau,\vec{y}) 
 \,\rangle
 & \stackrel{ \tau \ge -{\epsilon} / {k^{ }_\rmii{max}} }{=} &
 \int_\vec{k} e^{i\vec{k}\cdot(\vec{x-y})}_{ }
 \, 
 \frac{H^2}{ 2k^3} \bigl( 1 + k^2\tau^2 ) 
 \\[2mm] 
 & \equiv &
 \biggl( \frac{H}{2\pi} \biggr)^2_{ }
 \int_{k^{ }_\rmii{min}}^{k^{ }_\rmii{max}} 
 \! \frac{{\rm d}k}{k^2} \,
 \frac{\sin(k|\vec{x-y}|)}{|\vec{x-y}|}
 \bigl( 1 + k^2\tau^2 \bigr)
 \;. \hspace*{6mm} \la{hx_hy_via_k}
\ea
Here we have introduced co-moving momentum cutoffs, 
like in \fig\ref{fig:horizon}, as otherwise the 
integrals are not necessarily convergent.
The constraint on $\tau$ corresponds to that 
mentioned below \eq\nr{g_f_sum}, and will be assumed
implicitly in the following.
The equal-position limit $\vec{y}\to\vec{x}$
shows strong dependence on the momentum cutoffs,  
\be
  \langle\, 
 h^{ }_{>}(\tau,\vec{x}) \, 
 h^{ }_{>}(\tau,\vec{x}) 
 \,\rangle
 \; = \; 
 \biggl( \frac{H}{2\pi} \biggr)^2_{ }
 \biggl[ \, 
  \ln\biggl( \frac{k^{ }_\rmii{max}}{k^{ }_\rmii{min}} \biggr)
 \; + \; 
 \frac{\tau^2 ( k^{2}_\rmii{max} - k^{2}_\rmii{min} ) }{2}
 \, \biggr]
 \;. \la{hx_hy_tt}
\ee
As visible from \fig\ref{fig:horizon}, 
$k^{ }_\rmii{min}$ and $k^{ }_\rmii{max}$ can be mapped
onto the starting and ending points of horizon crossing. 

For a clear physical picture and as a crosscheck, it would be 
nice to evaluate \eq\nr{hx_hy} directly, without going via
momentum space, like in \eqs\nr{hx_hy_via_k} and \nr{hx_hy_tt}. 
We show here how this can be done.  
As a first step, we then 
need the noise autocorrelator, given by \eq\nr{noise_qm}. 

The functions $f^{ }_k$ appearing in \eq\nr{noise_qm} are
defined in \eq\nr{rho_Q}. Inserting
\be
 W'_k(\tau) =  -\frac{\epsilon}{\tau^2}\,
 \delta\biggl( k + \frac{\epsilon}{\tau}\biggr)
 \;, \quad
 W''_k(\tau) = \frac{2\epsilon}{\tau^3}\,
 \delta\biggl( k + \frac{\epsilon}{\tau}\biggr) 
 +\frac{\epsilon^2}{\tau^4}\,
 \delta'\biggl( k + \frac{\epsilon}{\tau}\biggr)
 \;, \la{W_k}
\ee
we can write 
\ba
 f^{ }_k(\tau^{ }_1) & = & 
 \frac{i H e^{-i k \tau^{ }_1}_{ }}{\sqrt{2k^3}}
 \biggl[
   \delta'\biggl( k + \frac{\epsilon}{\tau^{ }_1} \biggr)
   \,
   \frac{\epsilon^2(1 + i k \tau^{ }_1)}{\tau_1^4}
   + 
   \delta\biggl( k + \frac{\epsilon}{\tau^{ }_1} \biggr)
   \,
   \frac{\epsilon(4 + 4 i k \tau^{ }_1 - 2 k^2 \tau_1^2)}{\tau_1^3}
 \biggr] \hspace*{6mm}
 \\[2mm] 
 & = & 
 \frac{i H }{\sqrt{2k^3}}
 \biggl\{ \, 
 -\frac{{\rm d}}{{\rm d}\tau^{ }_1}
   \biggl[\, 
   \delta\biggl( k + \frac{\epsilon}{\tau^{ }_1} \biggr)
   \,
   \frac{\epsilon(1 + i k \tau^{ }_1)}{\tau_1^2}
   e^{-i k \tau^{ }_1}_{ }
   \,\biggr]
    + \,
   \delta\biggl( k + \frac{\epsilon}{\tau^{ }_1} \biggr)
   \,
   \frac{\epsilon(2 + 2 i k \tau^{ }_1 - k^2 \tau_1^2)}{\tau_1^3}
   e^{-i k \tau^{ }_1}_{ }
  \,\biggr\}
  \;.
\ea
In \eq\nr{noise_qm}, we pull the time derivatives in front of the
$\vec{k}$-integral. Then the integrand contains two Dirac-$\delta$'s, 
which can be expressed as 
\be
   \delta\biggl( k + \frac{\epsilon}{\tau^{ }_1} \biggr)
   \delta\biggl( k + \frac{\epsilon}{\tau^{ }_2} \biggr)
 = 
   \delta\biggl( \frac{\epsilon}{\tau^{ }_1}
              - \frac{\epsilon}{\tau^{ }_2} \biggr)
   \delta\biggl( k + \frac{\epsilon}{\tau^{ }_1} \biggr)
 = 
   \frac{\tau^{ }_1\tau^{ }_2}{\epsilon}\,
   \delta(\tau^{ }_1 - \tau^{ }_2)
   \delta\biggl( k + \frac{\epsilon}{\tau^{ }_1} \biggr)
 \;. 
\ee
Carrying out the angular and radial integrals, the latter
constrained to $k^{ }_\rmi{min} < k < k^{ }_\rmi{max}$
like in \eq\nr{hx_hy_via_k}, yields
\ba
 & & \hspace*{-1.5cm}
 \langle 0 | \, 
 \xiQ^{ }(\tau^{ }_1,\vec{z}) \, 
 \xiQ^{ }(\tau^{ }_2,\vec{v}) 
 \, | 0 \rangle 
 \nn[2mm]
 & = & 
 - \frac{H^2}{4\pi^2\epsilon}
   \biggl[\,
     (1+\epsilon^2_{ })\,
         \partial^{ }_{\tau_1}
         \partial^{ }_{\tau_2}
  - (2 + \epsilon^2_{ }+ i \epsilon^3_{ })\,
         \partial^{ }_{\tau_1} 
         \frac{1}{\tau^{ }_1} 
  - (2 + \epsilon^2_{ }- i \epsilon^3_{ })\, 
         \partial^{ }_{\tau_2} 
         \frac{1}{\tau^{ }_2}
 \nn[2mm] 
 & & 
  + \, (4+\epsilon^4)\, 
         \frac{1}{\tau^{ }_1\tau^{ }_2} 
   \,\biggr] 
   \frac{ 
    \delta(\tau^{ }_1 - \tau^{ }_2) }
        { |\vec{z-v}| }
   \sin\biggl( \frac{\epsilon |\vec{z-v}|}{\tau^{ }_1} \biggr)
   \theta\biggl(\,
          -\frac{\epsilon}{k^{ }_\rmii{min}} 
           \le \tau^{ }_1 \le 
          -\frac{\epsilon}{k^{ }_\rmii{max}}
         \,\biggr)
 \;, \la{noise_final}
\ea
where we denoted 
$\theta(\mbox{true}) \equiv 1$,
$\theta(\mbox{false}) \equiv 0$.
We stress that this representation is 
meaningful only as a part of an 
integrand containing test functions depending
on $\tau^{ }_1$ and $\tau^{ }_2$,
like \eq\nr{hx_hy}. The derivatives need then 
to be partially integrated, so that they act
on the test functions.  

We now return to \eq\nr{hx_hy} and insert \eq\nr{noise_final} there. 
Carrying out the partial integrations, and noting that boundary terms do
not contribute, we obtain
\ba
 \langle\, 
 h^{ }_{>}(\tau,\vec{x}) \, 
 h^{ }_{>}(\tau,\vec{y}) 
 \,\rangle
 & = & 
 - \frac{H^2}{4\pi^2\epsilon}
   \int_{\vec{z},\vec{v}}
   \int_{-\frac{\epsilon}{k_\rmii{min}}}
       ^{-\frac{\epsilon}{k_\rmii{max}}}
       \! {\rm d}\tau^{ }_1 \, 
   \biggl[\, 
       (1 + \epsilon^2) 
       \partial^{ }_{\tau_1} G^{ }_{|\vec{x-z}|}(\tau,\tau^{ }_1)\,
       \partial^{ }_{\tau_1} G^{ }_{|\vec{y-v}|}(\tau,\tau^{ }_1)
 \nn
 && \; + \, 
     \frac{2+\epsilon^2 + i \epsilon^3}{\tau^{ }_1} 
       \partial^{ }_{\tau_1} G^{ }_{|\vec{x-z}|}(\tau,\tau^{ }_1)\,
                             G^{ }_{|\vec{y-v}|}(\tau,\tau^{ }_1)
 \nn
 && \; + \, 
     \frac{2+\epsilon^2 - i \epsilon^3}{\tau^{ }_1} 
                             G^{ }_{|\vec{x-z}|}(\tau,\tau^{ }_1)\,
       \partial^{ }_{\tau_1} G^{ }_{|\vec{y-v}|}(\tau,\tau^{ }_1)
 \nn
 && \; + \, 
     \frac{4+\epsilon^4}{\tau^{2}_1} 
                             G^{ }_{|\vec{x-z}|}(\tau,\tau^{ }_1)\,
                             G^{ }_{|\vec{y-v}|}(\tau,\tau^{ }_1)
 \, \biggr]  
 \, \frac{1}
         {|\vec{z-v}|}
    \sin\biggl( \frac{\epsilon |\vec{z-v}|}{\tau^{ }_1} \biggr)
 \;. \hspace*{6mm} \la{spatial_1}
\ea

The spatial integration range pertinent to \eq\nr{spatial_1} 
is somewhat complicated. In order to simplify the task, we set 
$
 \vec{y} \to \vec{x} \to \vec{0}
$
(in view of translational invariance, 
the last step is inconsequential). 
Substituting also $\vec{z}\leftrightarrow\vec{v}$
in the middle term, we thereby obtain
\ba
 \langle\, 
 h^{ }_{>}(\tau,\vec{x}) \, 
 h^{ }_{>}(\tau,\vec{x}) 
 \,\rangle
 & = & 
 - \frac{H^2}{4\pi^2\epsilon}
   \int_{-\frac{\epsilon}{k_\rmii{min}}}
       ^{-\frac{\epsilon}{k_\rmii{max}}}
       \! {\rm d}\tau^{ }_1 \, 
   \int_{\vec{z},\vec{v}}
   \biggl[\, 
       (1 + \epsilon^2) 
       \partial^{ }_{\tau_1} G^{ }_{z}(\tau,\tau^{ }_1)\,
       \partial^{ }_{\tau_1} G^{ }_{v}(\tau,\tau^{ }_1)
 \nn
 && 
     \; + \, 
     \frac{2(2+\epsilon^2)}{\tau^{ }_1} 
       \partial^{ }_{\tau_1} G^{ }_{z}(\tau,\tau^{ }_1)\,
                             G^{ }_{v}(\tau,\tau^{ }_1)
 \nn
 &&
     \; + \, 
     \frac{4+\epsilon^4}{\tau^{2}_1} 
                             G^{ }_{z}(\tau,\tau^{ }_1)\,
                             G^{ }_{v}(\tau,\tau^{ }_1)
 \, \biggr] 
 \, \frac{1}
         {|\vec{z-v}|}
    \sin\biggl( \frac{\epsilon |\vec{z-v}|}{\tau^{ }_1} \biggr)
 \;. \hspace*{6mm} \la{spatial_2}
\ea

As a next step, we carry out the spatial integrals. 
To this aim, consider the general structure
\ba
 I(\tau,\tau^{ }_1,\tau^{ }_2) & \equiv & 
 \int_{\vec{z},\vec{v}} 
 G^{ }_z(\tau,\tau^{ }_1)\, G^{ }_v(\tau,\tau^{ }_2)
 \, \alpha(\tau^{ }_1,\tau^{ }_2,|\vec{z-v}|)
 \nn 
 & = & 
 \int_0^\infty \! {\rm d}w\, \alpha(\tau^{ }_1,\tau^{ }_2,w)
 \int_{\vec{z},\vec{v}} 
 G^{ }_z(\tau,\tau^{ }_1)\, G^{ }_v(\tau,\tau^{ }_2)
 \, \delta(w - |\vec{z-v}|)
 \nn 
 & = & 
 8 \pi^2 
 \int_0^\infty \! {\rm d}w\, w \, \alpha(\tau^{ }_1,\tau^{ }_2,w)
 \int_0^\infty \! {\rm d}z\, z 
 \int_{|w-z|}^{w+z} \! {\rm d}v\, v \, 
 G^{ }_z(\tau,\tau^{ }_1)\, G^{ }_v(\tau,\tau^{ }_2)
 \;, \la{def_I}
\ea
where in the last step we performed the angular integral. 
The Green's functions are 
inserted from \eqs\nr{g_x_lc} and \nr{g_x_tail}. There are
four different structures. 
The lightcone-lightcone part yields 
\ba
 && \hspace*{-2.5cm} 
 \int_0^\infty \!\!\! {\rm d}z\, z 
 \int_{|w-z|}^{w+z} \!\!\! {\rm d}v\, v \, 
 \delta(\tau - \tau^{ }_1 - z) \,
 \delta(\tau - \tau^{ }_2 - v)
 \nn[2mm] 
 & \stackrel{\tau^{ }_1 \ge \tau^{ }_2}{=} &  
 \theta(\tau^{ }_1 - \tau^{ }_2 \le w \le 2\tau - \tau^{ }_1 - \tau^{ }_2)\,
 \, 
 (\tau - \tau^{ }_1)(\tau - \tau^{ }_2)
 \;, \hspace*{6mm} \la{l1l2}
\ea
where the restriction to $\tau^{ }_1 \ge \tau^{ }_2$ is just 
a convenience (in the end $\tau^{ }_2\to \tau^{ }_1$).
The tail-lightcone and lightcone-tail contributions can be expressed as
\ba
 && \hspace*{-1.5cm} 
 \int_0^\infty \!\!\! {\rm d}z\, z 
 \int_{|w-z|}^{w+z} \!\!\! {\rm d}v\, v \, 
 \theta(\tau - \tau^{ }_1 - z) \,
 \delta(\tau - \tau^{ }_2 - v)
 \nn[2mm] 
 & \stackrel{\tau^{ }_1 \ge \tau^{ }_2}{=} &  
 \theta(\tau^{ }_1 - \tau^{ }_2 \le w \le 2\tau - \tau^{ }_1 - \tau^{ }_2)\,
 \, 
 \frac{(\tau - \tau^{ }_2)
 (2\tau - \tau^{ }_1 - \tau^{ }_2 - w)  
 (\tau^{ }_2 - \tau^{ }_1 + w)}{2}
 \;, \hspace*{6mm} \\
 && \hspace*{-1.5cm} 
 \int_0^\infty \!\!\! {\rm d}z\, z 
 \int_{|w-z|}^{w+z} \!\!\! {\rm d}v\, v \, 
 \delta(\tau - \tau^{ }_1 - z) \,
 \theta(\tau - \tau^{ }_2 - v)
 \nn[2mm] 
 & \stackrel{\tau^{ }_1 \ge \tau^{ }_2}{=} &  
 \theta(0 \le w \le \tau^{ }_1 - \tau^{ }_2) 
 \, 
 2(\tau - \tau^{ }_1)^2 w 
 \nn[2mm] 
 & + &  
 \theta(\tau^{ }_1 - \tau^{ }_2 \le w \le 2\tau - \tau^{ }_1 - \tau^{ }_2)\,
 \, 
 \frac{(\tau - \tau^{ }_1)
 (2\tau - \tau^{ }_1 - \tau^{ }_2 - w)  
 (\tau^{ }_1 - \tau^{ }_2 + w)}{2}
 \;. \hspace*{6mm} 
\ea
Finally, the tail-tail part evaluates to 
\ba
 && \hspace*{-1.5cm} 
 \int_0^\infty \!\!\! {\rm d}z\, z 
 \int_{|w-z|}^{w+z} \!\!\! {\rm d}v\, v \, 
 \theta(\tau - \tau^{ }_1 - z) \,
 \theta(\tau - \tau^{ }_2 - v)
 \nn[2mm] 
 & \stackrel{\tau^{ }_1 \ge \tau^{ }_2}{=} &  
 \theta(0 \le w \le \tau^{ }_1 - \tau^{ }_2) 
 \, 
 \frac{2(\tau - \tau^{ }_1)^3 w}{3} 
 \nn[2mm] 
 & + &  
 \theta(\tau^{ }_1 - \tau^{ }_2 \le w \le 2\tau - \tau^{ }_1 - \tau^{ }_2)\,
 \; 
 \frac{(w+\tau^{ }_1 + \tau^{ }_2 - 2 \tau)^2
 [w^2 + 2 w(2\tau - \tau^{ }_1 - \tau^{ }_2) - 
 3(\tau^{ }_1 - \tau^{ }_2)^2]}{24}
 \;. \hspace*{6mm} \la{t1t2}
\ea

We now insert \eqs\nr{l1l2}--\nr{t1t2} into \eq\nr{def_I}, together
with the function $\alpha$ from \eq\nr{spatial_2}. 
The integral over $w = |\vec{z-v}|$ can be carried out. 
For instance, 
\ba
 I(\tau,\tau^{ }_1,\tau^{ }_2) \bigg|^{
    G \to g^{ }_\rmii{tail}
 }_
 { 
   \alpha \to \frac{1}{w} \sin ({\epsilon w} / {\tilde \tau^{ }_1})
 }
 & = & 
 \frac{\tilde \tau_1^3}{\epsilon^5 \tau_1^2 \tau_2^2}
 \biggl\{
  \tilde \tau^{ }_1 
  \sin\biggl[ \frac{\epsilon(\tau - \tau^{ }_1)}
                   {\tilde \tau^{ }_1} \biggr] 
 - \epsilon (\tau - \tau^{ }_1 )
  \cos\biggl[ \frac{\epsilon(\tau - \tau^{ }_1)}
                   {\tilde \tau^{ }_1} \biggr] 
 \biggr\}
 \nn[3mm]
 & \times & 
 \biggl\{
  \tilde \tau^{ }_1 
  \sin\biggl[ \frac{\epsilon(\tau - \tau^{ }_2)}
                   {\tilde \tau^{ }_1} \biggr] 
 - \epsilon (\tau - \tau^{ }_2 )
  \cos\biggl[ \frac{\epsilon(\tau - \tau^{ }_2)}
                   {\tilde \tau^{ }_1} \biggr] 
 \biggr\}
 \;. 
\ea
Subsequently we operate
with the derivatives on the time arguments $\tau^{ }_1$ and 
$\tau^{ }_2$, as dictated by \eq\nr{spatial_2}. 
At this stage it is important that ${\tilde \tau^{ }_1}$, 
originating from the argument of 
$\sin(\epsilon w / \tilde \tau^{ }_1)$, 
is {\em not} differentiated, 
which is why we have given it a different name.  
After the differentiation, we set 
$\{ \tilde \tau^{ }_1,\tau^{ }_2 \} \to \tau^{ }_1$.
For instance, 
\ba
 \langle\, 
 h^{ }_{>}(\tau,\vec{x}) \, 
 h^{ }_{>}(\tau,\vec{x}) 
 \,\rangle^{ }_\rmi{tail-tail}
 & = & 
 - \frac{H^2}{4\pi^2}
   \int_{-\frac{\epsilon}{k_\rmii{min}}}
       ^{-\frac{\epsilon}{k_\rmii{max}}}
       \! {\rm d}\tau^{ }_1 \, 
 \biggl\{
   \frac{1}{\tau^{ }_1}
 + \frac{\tau}{\epsilon\tau^2_1}
   \biggl[ 
     \sin\frac{2\epsilon(\tau - \tau^{ }_1)}{\tau^{ }_1}
     - 2\epsilon
   \biggr] 
 \nn[2mm] 
 && 
 \; - \, \frac{\tau^2}{2\epsilon^2\tau^3_1}
   \biggl[ 
     \cos\frac{2\epsilon(\tau - \tau^{ }_1)}{\tau^{ }_1}
     + 2\epsilon
     \sin\frac{2\epsilon(\tau - \tau^{ }_1)}{\tau^{ }_1}
     - 1 - 2\epsilon^2 
   \biggr] 
 \biggr\}
 \;. \la{x_penul}
\ea
Adding all terms, a significant cancellation takes place, 
yielding finally 
\be
 \langle\, 
 h^{ }_{>}(\tau,\vec{x}) \, 
 h^{ }_{>}(\tau,\vec{x}) 
 \,\rangle
  = 
 - \frac{H^2}{4\pi^2}
   \int_{-\frac{\epsilon}{k_\rmii{min}}}
       ^{-\frac{\epsilon}{k_\rmii{max}}}
       \! {\rm d}\tau^{ }_1 \, 
 \biggl(
   \frac{1}{\tau^{ }_1}
  + 
   \frac{\epsilon^2\tau^2}{\tau_1^3}
 \biggr) 
 \;. \la{x_final}
\ee
The integral is easily done (though we need to note that $\tau^{ }_1 < 0$
so that the parentheses are negative, and watch out with signs). 
Thereby \eq\nr{hx_hy_tt} is reproduced. 
The phenomenologically relevant term, 
ending up in \eq\nr{P_T}, originates from 
the $1/\tau^{ }_1$ in \eq\nr{x_final}, 
which in turn comes from the tail-tail contribution 
in \eq\nr{x_penul}. 

\end{document}